\documentclass[a4paper,12pt]{article}

\usepackage{natbib}
\usepackage[english]{babel}

\usepackage{booktabs,dcolumn}
\usepackage{psfrag,graphicx}
\usepackage{eso-pic,graphicx}
\usepackage{amsmath}
\usepackage{graphicx}
\usepackage{pdflscape}
\usepackage{longtable}
\usepackage{epstopdf}
\usepackage{appendix}
\usepackage{dashbox}
\usepackage{algorithm,algcompatible,amsmath}
\algnewcommand\INPUT{\item[\textbf{Input:}]}%
\algnewcommand\OUTPUT{\item[\textbf{Output:}]}%

\newcolumntype{z}[1]{D{.}{.}{#1}}

\date{}

\begin{document}

\title{
\begin{center}
{\Large \bf Modelling uncertainty in financial tail risk: a forecast combination and weighted quantile approach} \end{center}
}

\author{Giuseppe Storti$^{1}$\footnote{Corresponding author. Address correspondence to: Giuseppe Storti, Università degli studi di Salerno, Dipartimento di Scienze Economiche e Statistiche (DISES), Via Giovanni Paolo II, 132, 84084 Fisciano (SA), Italy. Email: storti@unisa.it.}, Chao Wang$^{2}$
\\
$^{1}$ Department of Economics and Statistics, University of Salerno, \\
$^{2}$ Discipline of Business Analytics, The University of Sydney}
\date{} \maketitle

\begin{abstract}
\noindent

A novel forecast combination and weighted quantile based tail-risk forecasting framework is proposed, aiming to reduce the impact of modelling uncertainty in tail-risk forecasting. The proposed approach is based on a two-step estimation procedure. The first step involves the combination of Value-at-Risk (VaR) forecasts at a grid of quantile levels. A range of parametric and semi-parametric models is selected as the model universe in the forecast combination procedure. The quantile forecast combination weights are estimated by optimizing the quantile loss. In the second step, the Expected Shortfall (ES) is computed as a weighted average of combined quantiles. The quantiles weighting structure for ES forecasting is determined by minimizing a strictly consistent joint VaR and ES loss function of the Fissler-Ziegel class. The proposed framework is applied to six stock market indices and its forecasting performance is compared to each individual model in the universe, a simple average approach and a weighted quantile approach. The forecasting results  support the proposed framework.

%A novel forecast combination and weighted quantile based tail risk forecasting framework is proposed, aiming to reduce the impact of modelling uncertainty in financial tail risk forecasting. The proposed approach is based on a two-step estimation procedure. The first step involves the combination of Value-at-Risk (VaR) forecasts at a grid of different quantile levels. A range of parametric and semi-parametric models is selected as the model universe which is incorporated in the forecast combination procedure. The quantile forecast combination weights are estimated by optimizing the quantile loss. In the second step, the Expected Shortfall (ES) is computed as a weighted average of combined quantiles. The quantiles weighting structure used to generate the ES forecast is determined by minimizing a strictly consistent joint VaR and ES loss function of the Fissler-Ziegel class. The proposed framework is applied to six stock market indices and its forecasting performance is compared to each individual model in the model universe and a simple average approach. The forecasting results based on a number of evaluations support the proposed framework.

\vspace{0.5cm}


\end{abstract}

\newpage
\pagenumbering{arabic}

{\centering
\section{Introduction}\label{introduction_sec}
\par
}
\noindent

Since the introduction by J.P. Morgan in the RiskMetrics model in 1993, Value-at-Risk (VaR) has been widely employed by financial institutions and corporations around the world to assist their decision making in relation to capital allocation and risk management. VaR is a quantitative tool to measure and control financial risk and represents the market risk as one number. VaR has become a standard measurement for capital allocation and risk management. Let $\mathcal{I}_t$ be the information available at time $t$ and
\[
F_{t}(r)=Pr(r_{t}\leq r | \mathcal{I}_{t-1})
\]
be the Cumulative Distribution Function (CDF) of return $r_{t}$ conditional on $\mathcal{I}_{t-1}$. We assume that $F_{t}(.)$ is strictly increasing and continuous on the real line $\Re$. Under this assumption, the $\alpha$ level VaR (quantile) at time $t$ can be defined as:
\begin{equation} \label{var_def}
Q_{t,\alpha}=F^{-1}_{t}(\alpha), \qquad 0 <\alpha <1. \nonumber
\end{equation}

However, VaR has been subject to criticism because it cannot measure the expected loss for violations and is not mathematically coherent, in that it can favor non-diversification. Expected Shortfall (ES), proposed by \cite{artzner1997} and \cite{artzner1999}, gives the expected loss, conditional on returns exceeding a VaR threshold, and is a coherent measure; thus, in recent years it has become more widely employed for tail risk measurement and is now favored by the Basel Committee on Banking Supervision. Within the same framework as above, the $\alpha$ level ES can be shown to be equal to the tail conditional expectation of $r_t$ \citep[see][among others]{AceTas2002}:
\begin{equation}
    ES_{t,\alpha}=E(r_t|r_t\leq Q_{t,\alpha}, \mathcal{I}_{t-1}).
\label{e:ESdef}
\end{equation}

%Within the same framework, the $\alpha$ level ES can be shown \citep[see][among others]{AceTas2002} to be equal to the tail conditional expectation of $r_t$.

The Basel III Accord, which was implemented in 2019, places new emphasis on ES. Its recommendations for market risk management are illustrated in the 2019 document \textit{Minimum Capital Requirements for Market Risk} that says: ``ES must be computed on a daily basis for the bank-wide internal models to determine market risk capital requirements. ES must also be computed on a daily basis for each trading desk that uses the internal models approach (IMA).'' (\cite{BIS2019}, p. 89). According to the same document, when calculating ES banks must refer to the average loss in the tail below the 2.5\% quantile level. Therefore, in the empirical application of our paper, we focus on one-step-ahead tail risk forecasting at the $\alpha=2.5\%$ quantile level. In order to simplify notation, in the remainder, unless differently specified, the following notational conventions are adopted: $\text{ES}_{t,\alpha}\equiv ES_{t}$ and $Q_{t,\alpha} \equiv Q_t$, where $\alpha$ denotes the target 2.5\% level for the estimation of VaR and ES.

Forecasts of VaR and ES can be generated through a variety of different models. Some of these, such as completely specified GARCH models, are fully parametric since they rely on the exact  specification of the conditional distribution of returns and of the volatility dynamics. Differently, semi-parametric approaches require specific assumptions on the risk dynamics but without a return distribution assumption.
Semi-parametric models can be applied to generate forecasts of VaR alone, that is the case of the conditional autoregressive VaR (CAViaR) models proposed by \cite{caviar}, or joint forecasts of the pair (VaR, ES). A joint semi-parametric model that directly estimates both VaR and ES, referred to here as the ES-CAViaR model, is proposed by \cite{tayl2017}.  Through incorporating an Asymmetric Laplace (AL) distribution with a time-varying scale, a quasi-likelihood can be built to enable the joint estimation of the conditional VaR and conditional ES in this framework.

\cite{Fissler2016} develop a family of joint loss functions (or ``scoring rules'') for the associated VaR and ES series that are strictly consistent for the pair (VaR, ES), that is, they are uniquely minimized by the true VaR and ES series. Applying specific choices of functions in the class of joint loss functions of \cite{Fissler2016}, it can be shown that such a loss function is exactly the same as the negative of the AL log-likelihood function presented in \cite{tayl2017}. \cite{pattonetal2019} propose new dynamic models for VaR and ES, through adopting the generalized autoregressive score (GAS) framework (\citealt{crealetal2013} and \citealt{harvey2013}) and utilizing the loss functions in \cite{Fissler2016}.

Alternatively, a variety of semi-parametric approaches to the prediction of VaR and ES can be obtained by combining Quasi Maximum Likelihood (QML) estimation of the volatility coefficients with some non-parametric estimator of the error quantiles. Widely diffused and effective solutions rely on extreme value theory results, such as in the peaks-over-threshold approach \citep{gilli2006application}.

\cite{storti2020nonparametric} have recently proposed a new ES estimation and forecasting framework, referred to as the Weighted Quantile (WQ) approach, where the ES is modelled as weighted average of tail quantiles. The quantiles are produced from the CAViaR model of \cite{caviar} by grid search of a range of equally spaced quantile levels below the target VaR level, i.e., 2.5\%. An advantage of this approach is that it sensibly reduces the impact of model uncertainty in the prediction of ES, that is modelled according to its natural definition as an average of tail quantiles. However, the specification of the optimal dynamic model for each quantile level is still subject to uncertainty. In order to limit the impact of the overall model uncertainty on the generation of joint (VaR, ES) forecasts, the WQ framework could actually be extended by replacing forecasts of tail quantiles from a single CAViaR model, as in \cite{storti2020nonparametric}, with forecast combinations from an ensemble of different models, of a possibly heterogeneous nature, that includes parametric as well as semi-parametric models. Then the ES forecasts could be generated as weighted averages of ``combined'' VaR predictors at different levels, employing the WQ framework.

The main motivation for this extension of the WQ framework relies on the consideration that, given the values of conditional tail quantiles, the ES is theoretically defined as the expectation of these quantiles. So, it can be immediately recognized that most of the modelling uncertainty is related to the modelling of tail quantiles, which will be addressed in this paper. The only residual uncertainty affecting ES estimation is potentially related to the identification of the grid of tail quantiles. However, as extensively discussed in \cite{storti2020nonparametric}, the ES estimates obtained through the WQ approach are not particularly sensitive to the selection of these hyper-parameters.

%the available computing power,  even on standard personal computers, easily allows

%allowing different quantiles to be estimated with different models (not limited to CAViaR).

%since the optimal model on modelling the quantile on each quantile level is uncertain,

%In this paper, for each quantile level in the WQ approach, we propose incorporating the forecasting combination approach to produce the quantile estimates and forecasts which are potentially more robust than the ones from individual models.
%Then the ES forecasts are produced with a set of averaged VaR predictors at different levels, employing the WQ framework as in \cite{storti2020nonparametric}.

Aim of this paper is then to propose a novel approach to forecast VaR and ES based on Forecast Combination and Weighted Quantile (FC-WQ) techniques. The proposed framework can be treated as a generalization of the WQ framework proposed by \cite{storti2020nonparametric} and its effectiveness is investigated through applications to stock market data, where direct comparisons between the FC-WQ and WQ approaches are conducted.

The paper is structured as follows. A review of strictly consistent scoring functions for joint estimation of VaR and ES is conducted in Section \ref{score_functions}. Section \ref{methodology_sec} presents the the proposed approach and discusses its technical implementation details. The selected model universe for forecast combination is shown in Section \ref{model_universe}. The results of an empirical application to real stock market data are presented and discussed in Section \ref{data_empirical_section}. Finally, Section \ref{conclusion} concludes.

{\centering
\section{Strictly consistent scoring functions for joint estimation of VaR and ES}\label{score_functions}
\par
}

% Under the assumption $F_{t}(.)$ is strictly increasing and continuous on the real line $\Re$, the one-step-ahead $\alpha$ level Value at Risk at time $t$ can be defined as
% \[
% Q_{t,\alpha}=F^{-1}_{t}(\alpha)\qquad 0 <\alpha <1.
% \]
% Within the same framework, the one-step-ahead $\alpha$ level Expected Shortfall can be shown \citep[see][among others]{AceTas2002} to be equal to the tail conditional expectation of $r_t$
% \begin{equation}
%     ES_{t,\alpha}=E(r_t|r_t\leq Q_{t,\alpha}, \mathcal{I}_{t-1}).
% \label{e:ESdef}
% \end{equation}
%  In order to simplify notation, in the paper the following notational conventions are adopted: $\text{ES}_{t,\alpha}\equiv ES_{t}$ and $Q_{t,\alpha} \equiv Q_t$, where $\alpha$ denotes the target level for the estimation of VaR and ES (in the empirical application we focus on $\alpha=0.025$).

From the main definition of ES provided in Equation (\ref{e:ESdef}), it follows that that $\text{ES}_{t}$ is related to $F_{t}(.)$ by the following integral
\begin{equation}
    ES_{t}=\frac{1}{F_t(Q_{t})}\int_{-\infty}^{Q_{t}}rdF_{t}(r)=\frac{1}{\alpha}\int_{-\infty}^{Q_{t}}rdF_{t}(r),
\label{e:int1}
\end{equation}
%\int_{-\infty}^{Q_{t,\alpha}}rf_{t}(r)dr
%where $f_{t}(r)=\partial F_t(r) / \partial r$ is the conditional Probability Density Function (PDF) of $r_t$.
that, after a simple change of variable, can be rewritten as
\begin{equation}
    ES_{t}=\frac{1}{\alpha}\int_{0}^{\alpha}Q_{t,p}dp.
\label{e:int2}
\end{equation}

In the literature, several alternative parameterizations of $ES_t$ and $VaR_t$ have been proposed. The involved parameters can be consistently estimated from real data by minimizing appropriately chosen strictly consistent scoring functions. If the interest is solely in the estimation of VaR, the unknown coefficients in the dynamic specification of $Q_t$ can be estimated by quantile regression mimizing the expected quantile loss function
\[
QL(r_t,Q_t;\alpha)=(\alpha-I_{t})(r_t-Q_t)
\]
with $I_t=I(r_t<Q_t)$, where $I(A)$ is the indicator function taking value 1 if event $A$ occurs and 0 otherwise, for $t=1,\ldots,N$.

\cite{koenkermachado1999} show that the quantile regression estimator is equivalent to a maximum likelihood estimator when assuming that the data are conditionally distributed as an AL with a mode at the quantile of interest. If $r_t$ is the return on day $t$ and $Pr(r_t < Q_t | \mathcal{I}_{t-1}) = \alpha$, then the parameters in the model for $Q_t$ can be estimated by maximizing a quasi-likelihood based on:
$$ p(r_t| \mathcal{I}_{t-1}) = \frac{\alpha (1-\alpha)}{\sigma}  \exp \left( \frac{-(r_t-Q_t)(\alpha - I(r_t < Q_t) ) } {\sigma } \right), $$
for $t=1,\ldots,N$ and where $\sigma$ is a scale parameter.

\cite{tayl2017} extends this result to incorporate the associated ES quantity into the likelihood expression, noting a link between $\text{ES}_t$ and a dynamic $\sigma_t$, resulting in the conditional density function:
\begin{eqnarray} \label{es_var_likelihood}
p(r_t| \mathcal{I}_{t-1}) = \frac{(\alpha - 1)}{ES_t} \exp \left( \frac{(r_t-Q_t)(\alpha - I(r_t < Q_t)) }{\alpha ES_t}  \right).
\end{eqnarray}
This allows a likelihood function to be built and maximised, given model expressions for $(Q_t, \, ES_t)$. In Equation (\ref{es_var_likelihood}),  $r_{t}$ is the daily return, $Q_{t}$ and $\text{ES}_{t}$ denote the $\alpha$ target level VaR and ES on day $t$. \cite{tayl2017} notes that the negative logarithm of the resulting likelihood function is strictly consistent
for $(Q_t,\, ES_t)$ considered jointly, i.e., it fits into the class of jointly consistent scoring functions for VaR and ES developed by \cite{Fissler2016}.

Members of this family are strictly consistent for $(Q_t,ES_t)$, i.e., their expectations are uniquely minimized by the true VaR and ES series. The general form of this functional family is:
\begin{eqnarray}
S_t(r_t, Q_t, ES_t) &=& (I_t -\alpha)G_1(Q_t) - I_tG_1(r_t) +  G_2(ES_t)\left(ES_t-Q_t + \frac{I_t}{\alpha}(Q_t-r_t)\right) \nonumber\\
                      &-& H(ES_t) + a(r_t) \, ,
                      \label{e:fzloss}
\end{eqnarray}
%$I_t=1$ if $r_t<Q_t$ and 0 otherwise, for $t=1,\ldots,N$,
where $G_1(.)$ is increasing, $G_2(.)$ is strictly increasing and strictly convex,
$G_2 = H^{'}$ and $\lim_{x\to -\infty} G_2(x) = 0$ and $a(\cdot)$ is a real-valued integrable function.

As discussed in \cite{tayl2017}, assuming $r_t$ to have zero mean, making the choices: $G_1(x) =0$,
$G_2(x) = -1/x$, $H(x)= -\text{log}(-x)$ and  $a= 1-\text{log} (1-\alpha)$, which satisfy the required criteria, returns the scoring function:
\begin{eqnarray}\label{es_caviar_log_score}
S_t(r_t, Q_t, ES_t) = -\text{log} \left( \frac{\alpha-1}{\text{ES}_t} \right) - {\frac{(r_t-Q_t)(\alpha-I(r_t < Q_t))}{\alpha \text{ES}_t}}.
\end{eqnarray}
%where the aggregated loss is indicates as $\mathbf{S} = \sum_{t=1}^N S_t$.
\cite{tayl2017} refers to Equation (\ref{es_caviar_log_score}) as the AL log score, which is a strictly consistent scoring function whose expectation is jointly minimized by the true VaR and ES series. The negative of Equation (\ref{es_caviar_log_score}) equals to the log of Equation (\ref{es_var_likelihood}) and can be treated as the AL log-likelihood.

{\centering
\section{Proposed framework}\label{methodology_sec}
\par
}

\subsection{The weighted quantile framework}\label{wq_review}
\noindent
In the Weighted Quantile (WQ) framework, $\alpha$ level ES forecasts from a given model are generated in two steps. In step 1, given a grid of quantile levels of size $M$ $\alpha_j\leq \alpha$, $j=1,\ldots,M$, with $0<\alpha_1<\alpha_2<\ldots<\alpha_M=\alpha$, an ensemble of $M$ VaR forecasts ($\widehat{Q}_{t}^{(\alpha_j)}$) is produced. In their empirical application \cite{storti2020nonparametric} use CAViaR models for the modelling of tail quantiles. However, it is worth remarking that any model, parametric or semi-parametric, could be used to generate VaR forecasts for any $\alpha_j\leq \alpha$. In principle, even different models, parametric and semi-parametric, could be used to fit quantiles at different levels. This feature makes the WQ framework highly flexible and adaptive.

In step 2, ES forecasts are then computed as an affine function of the tail quantile forecasts at levels $\alpha_j\leq \alpha$ as in Equation (\ref{e:insES_wq}):
\begin{equation}
\widehat{ES}^{(\text{WQ})}_{t}= w_{0}+\sum_{j=1}^{M} w_{j} \widehat{Q}_{t}^{(\alpha_j)} \, ,
\label{e:insES_wq}
\end{equation}
%, borrowed from the mixed frequency adaptive data sampling framework \citep[see]{ghysels2007midas}
where the weights $w_{j}$, $j=1,\dots,M$, are generated by some flexible and parsimonious function, such as the Beta function. Namely, for $j=1,\ldots,M$, we have $w_j=w\left(\frac{j}{M};a,b\right)$ with
\begin{equation}
w(x;a,b)=\frac{x^{a-1}(1-x)^{b-1}\Gamma(a+b)}{\Gamma(a)\Gamma(b)}.
\label{e:betalag}
\end{equation}
The main reasons for adopting the Beta specification to model the weights behaviour in Equation (\ref{e:insES_wq}) are its parsimony, since it only depends on two parameters, and flexibility. However, as discussed in \cite{storti2020nonparametric}, other parameterizations are feasible and, in particular, for sufficiently low values of $M$, the weights $w_j$ can be easily estimated individually as ``free'' parameters.

The intercept $w_0$ is estimated along with the other parameters and allows to correct biases potentially arising from the left truncation in the chosen grid of tail quantiles, thus further increasing the flexibility of the WQ approach.

Given first stage VaR forecasts, the unknown coefficients in Equation (\ref{e:insES_wq}) are estimated by optimizing some strictly consistent scoring function such as the AL \citep{tayl2017} or some other strictly consistent scoring function in the Fissler-Ziegel class.

\subsection{A unified framework encompassing Forecast Combination and Weighted Quantile estimation}\label{proposed_model}
It is important to note that model uncertainty mainly affects step 1 of the WQ procedure, since the estimation formula used in step 2 naturally stems from the mathematical definition of ES as a function of the tail VaRs. In step 1 of the procedure, different models could result in being optimal for different quantile levels. On the other hand, in step 2, the ES is estimated applying its natural definition as expectation of the tail quantiles. The only residual source of modelling uncertainty is related to the selection of the grid of tail quantiles used for estimation in Equation (\ref{e:insES_wq}) and, in particular, to the choice of the lower bound $\alpha_1$ and  of the value of $M$.
The impact of the choice of the lower bound is, by construction, controlled by the intercept $w_0$ and by the data driven weighting structure of the ES estimator, as also documented empirically by \cite{storti2020nonparametric}.
Regarding the choice of $M$, given the high computing power routinely available even on standard personal computers, implementing the WQ approach with a high value of $M$, that virtually eliminates the discretization error, is not an issue and could be easily done while still keeping the computing time at reasonable levels. On the other hand, \cite{storti2020nonparametric} show, by simulations and applications to real data, that the WQ gives remarkably good performances even for values of $M$ as low as 3 and that negligible improvements, in terms of forecasting accuracy, are expected from an increase in $M$.

These considerations motivate a two-step approach based on a Forecast Combination and Weighted Quantile (FC-WQ) strategy, that is the main contribution of this paper. The technical details on the implementation of the two steps of the FC-WQ procedure are presented as below.

\noindent \textbf{Step 1}:
Assume that $n_{mod}$ different VaR forecasting methods are available and that each of them is fitted to generate series of VaR forecasts for a set of strictly increasing quantile orders $\boldsymbol{\alpha_M}=[0<\alpha_1,\ldots,\alpha_M=\alpha]$. To be consistent with the standard regulatory prescription, we will focus on the case $\alpha=2.5\%$.

Letting $N$ be the length of the in-sample window and $T>>N$ the length of the full-sample returns series, for each available model in our \emph{model universe} ($n_{mod}$ models in total), $M$ VaR forecasts series will be generated from each model:
\begin{equation}
    \widehat{Q}^{(\alpha_j)}_{N+1,i},\ldots, \widehat{Q}^{(\alpha_j)}_{N+H,i}
    \label{e:varforc}
\end{equation}
%If a rolling window forecasting scheme is adopted,
for $i=1,\ldots,n_{mod}$, $j=1,\ldots,M$ and where $H=T-N$ is the length of the out-of-sample period to be used for forecast evaluation. Overall, this will yield a total of $M \times n_{mod}$ series of VaR forecasts. For the $i$-th model and $j$-th quantile level, the generic $h$-th one-step-ahead forecast $\widehat{Q}^{(\alpha_j)}_{N+h,i}$ ($h=1,\ldots,H$) will be based on the model fitted to observations from $h$ to $N+h-1$. Forecast combination is then used as a technique for reducing model uncertainty in VaR forecasting, yielding the \emph{combined quantile predictor} that takes the general form
\begin{equation}
\widehat{Q}^{(C,\alpha_j)}_{t}=c_{0,j}+\sum_{i=1}^{n_{mod}}c_{i,j}\widehat{Q}^{(\alpha_j)}_{t,i}.
\label{e:combpred}
\end{equation}

% \begin{equation}
% \bar{Q}^{(\alpha_j)}_{t}= -\text{exp}(c_{0,j}+\sum_{i=1}^{n_{mod}}c_{i,j} \text{log}( |\widehat{Q}^{(\alpha_j)}_{t,i}|).
% \label{e:combpred}
% \end{equation}

Therefore, in step 1, VaR forecasts from $n_{mod}$ different models are combined to generate a set of ``combined'' VaR predictors at different $\alpha_j$ quantile levels. In order to overcome the potential quantile crossing problem, the monotonization method proposed by \citet{chernozhukovetal2010} is employed.

The framework in (\ref{e:combpred}) can be in principle extended to consider non-linear combination schemes. This possibility is however not investigated in this paper. The proposed framework is highly flexible and can employ any model that could produce VaR estimates and forecasts, such as GARCH \citep{bollerslev1986generalized}, CAViaR, etc. Details of the model universe will be presented in Section \ref{model_universe}.

\noindent \textbf{Step 2}:
%In step 1, through combining $n_{mod}$ individual models the combined quantile predictor $\widehat{Q}^{(C,\alpha_j)}_{t}$ is produced for each quantile level $\alpha_j$.
%In order to overcome the potential quantile crossing problem, the monotonization method proposed by \citet{chernozhukovetal2010} is employed.
In step 2, employing the WQ approach, the conditional ES at time $t$ is modelled as the weighted average of the combined quantiles from step 1:
\begin{equation}
\widehat{ES}^{(\text{FC-WQ})}_{t}= w_{0}+\sum_{j=1}^{M} w_{j} \widehat{Q}^{(C,\alpha_j)}_{t} \, ,
\label{e:insES}
\end{equation}
where the weights $w_{i}$, $i=1,\dots,M$, are generated by a Beta weight function as in Equation (\ref{e:betalag})\footnote{Following the implementation of WQ in \citep{storti2020nonparametric}, we set the number of grid points equal to $M+1$, so that the weight of the $\alpha_{M}$-quantile is not 0 by construction when using the Beta weight function to parameterize the weights pattern.}.

% Namely, for $j=1,\ldots,M$, we have $w_j=w\left(\frac{j}{M};a,b\right)$ with
% \begin{equation}
% w(x;a,b)=\frac{x^{a-1}(1-x)^{b-1}\Gamma(a+b)}{\Gamma(a)\Gamma(b)}.
% \label{e:betalag}
% \end{equation}

%\noindent
%\textbf{Remark 1.} In step 1, for each quantile level $\alpha_j$, the model estimation and forecast combination can be run in parallel.

%\noindent
%\textbf{Remark 1.}
The combination of joint forecasts of VaR and ES has so far received scarce attention in the literature. The only contribution in this field is, to the extent of our knowledge, given by the paper of \cite{taylor2020forecast} whose approach is however structurally different from the one that is taken in this paper. First, \cite{taylor2020forecast} optimizes an AL loss function to combine joint (VaR,ES) models at a given target level $\alpha$, while our approach uses richer information on quantile levels falling in the tail of the distribution below $\alpha$. Second, our model universe is composed of VaR models rather than of joint (VaR, ES) models.

The next section focuses on the estimation strategy followed to estimate the $c_i$, for $i=1,...,n_{mod}$, and the intercept term $c_0$ in step 1, and the ($w_0$,$a$,$b$) coefficients in step 2.

\subsection{Implementation of the FC-WQ predictor: estimation procedure}\label{two_step_estimation}

Next, we provide a detailed description of the estimation procedures implemented in the two steps of the proposed FC-WQ  framework.
%Although, for ease of explanation, we focus on the standard risk level $\alpha=2.5\%$, the method can be immediately extended to other values of $\alpha$.\\

\noindent \textbf{Estimation Step 1:}

Under step 1, given the target quantile level $\alpha= 2.5\%$, an equally spaced grid of quantile levels of size $M$ is selected,
\[
\boldsymbol{\alpha_M}=\left[\alpha_1,\alpha_2,\ldots, \alpha_{M} \right],
\]
where $\alpha_j=\alpha_{j-1}+\eta$, with  $\alpha_M=\alpha$ and $\eta=(\alpha_M-\alpha_1)/(M-1)$, for $j=2,\ldots,M$. The value of the lower bound $\alpha_1$ is selected as 0.005 and $M$ is selected as 3 and 5 in this paper, according to the findings in \cite{storti2020nonparametric}. For example, with $M=3$ and $\alpha=2.5\%$, fixing $\alpha_1=0.005$ we have $\eta=0.01$ and the grid of quantile levels as $\boldsymbol{\alpha_M}= \,[0.005, 0.015, 0.025]$.

%%%%%%%%%%% NEW FOOTNOTE %%%%%%%%%%%%%%%%%%%%%%%% 14/06/2021

%\footnote{Note that the
%last element of the sequence of weights generated from
%the Beta weight function (i.e. the one corresponding to the risk level $\alpha_M=\alpha$) is by construction equal to 0, except when $b = 1$. To address this issue, in practical implementation of the WQ-Beta estimator, we set the actual number of grid points equal to $M + 1$, so that the weight of the $\alpha_M$-quantile is not 0 by construction. Referring to the example in the text, to estimate ES at level $\alpha_M = 2.5\%$ as a weighted average of M = 3 tail quantiles of order $\alpha_i\leq\alpha_M$, setting $\alpha_1 = 0.005$, we use the equally spaced grid of $M + 1 = 4$ points $[0.005, 0.015, 0.025,0.035]$ where the last quantile of order $0.035$ will always have weight equal to 0 and so it will not enter the computation of the ES predictor. It is worth noting that this issue does not arise if the weights are directly estimated or if we use a weighting function different from the Beta that does not impose the last weight to be equal to 0 by construction.}
%%%%%%%%%%%%%%%%%%%%%%%%%%%%%%%%%%%%%%%%%%%%%%%%%%%%%%%%%%%%%%%%

Now, for each trial quantile level $\alpha_j \in \boldsymbol{\alpha_M}$, $n_{mod}$ individual models are employed to produce the ($N \times 1$) time series of conditional in-sample quantiles $\widehat{\mathbf{Q}}_{1:N, i}^{(\alpha_j)}$ and the 1st one-step-ahead quantile forecasts $\widehat{Q}_{N+1, i}^{(\alpha_j)}$, for $i=1,\ldots,n_{mod}$ and $j=1,\ldots,M$. The set of in-sample quantiles (produced from in-sample data from $t=1$ to $N$ or $1:N$) for $n_{mod}$ models and at all trial quantile levels, from $\alpha_1$ to $\alpha_M$, is collected in the ${N \times (M\times n_{mod})}$ array $\widehat{\mathbf{Q}}_{1:N}$.

%For a given $j$, the coefficients $c_{i,j}$, for $i=0,\ldots,M$, can be estimated by minimizing the quantile loss function
%\[
%\widebar{QL}=\sum_{h=1}^{H}QL(r_{N+h},\bar{Q}^{(\alpha_j)}_{N+h};\alpha_j)=
%(\alpha_j-I_{N+h,j})(r_{N+h}-\bar{Q}^{(\alpha_j)}_{N+h})
%\]
%where $I_{t,j}=I(r_t<\bar{Q}^{(\alpha_j)}_{t})$, for each quantile level $j=1,\ldots, M$.

%More specifically, as in \cite{giacominikomunjer2005}, the weights are estimated via optimizing the quantile loss function over the out-of-sample period. This procedure is separately performed for each quantile order $\alpha_j$ so that different combination weights are obtained for each VaR level.

%Analytically, the estimated coefficients for combining time $N+h$ forecasts are given by
%\begin{equation}
%\hat{\mathbf{c}}_{j}=\underset{\mathbf{c}_{j}}{ \arg \min}\,\, QL_{N+h}(\mathbf{c}_j;\alpha_j)
%\label{e:combpred_est}
%\end{equation}
%where $\mathbf{c}_j=(c_{0,j},c_{1,j},\ldots, c_{n_{mod},j})'$ and

%\begin{equation}\label{q_loss}
%QL_{N+h}(\mathbf{c}_j;\alpha_j)=\frac{1}{N} \sum_{k=1}^{N}\left(\alpha_j-I\left(r_{N+h-k}<\bar{Q}^{(\alpha_j)}_{N+h-k}\right)\right)\left(r_{N+h-k}<\bar{Q}^{(\alpha_j)}_{N+h-k}\right)  \,\,
%\label{e:combpred_target}
%\end{equation}
%for $j=1,\ldots,M$. The procedure is initialized with in-sample VaR estimates which are gradually replaced with out-of-sample VaR forecasts.

In our paper, a rolling window forecasting scheme is adopted, with $T$ as the length of the full-sample returns series. Therefore, for each step in the $H=T-N$ out-of-sample periods (still for each trial quantile level $\alpha_j \in \boldsymbol{\alpha_M}$), $n_{mod}$ individual models are employed to produce the time series of one-step-ahead quantile forecasts $\widehat{Q}^{(\alpha_j)}_{N+1,i},\ldots, \widehat{Q}^{(\alpha_j)}_{N+H,i}$, for $i=1,\ldots,n_{mod}$ and $j=1,\ldots,M$. This set of out-of-sample quantiles for $n_{mod}$ models at all trial quantile levels is collected in the ${H \times (M\times n_{mod})}$ array $\widehat{\mathbf{Q}}_{(N+1):(N+H)}$, here $N+H=T$.

Concatenating the two matrices $\widehat{\mathbf{Q}}_{1:N}$ (including in-sample quantile estimates) and $\widehat{\mathbf{Q}}_{(N+1):(N+H)}$ (including one-step-ahead out-of-sample quantile forecasts) produces our \emph{quantile universe} $\widehat{\mathbf{Q}}^{(U)}_{1:(N+H)}$ which is an array of size $T \times (M \times n_{mod})$. The component of $\widehat{\mathbf{Q}}^{(U)}_{1:(N+H)}$ including in-sample VaR estimates and out-of-sample VaR forecasts at quantile level $\alpha_j$ will be denoted as $\widehat{\mathbf{Q}}^{(U,\alpha_j)}_{1:(N+H)}$, that is a $T \times n_{mod}$ matrix.

The $\widehat{\mathbf{Q}}^{(U,\alpha_j)}_{1:(N+H)}$ matrix and the series of returns $\mathbf{r}_{1:(N+H)}$ are then given as input to the estimation procedure for VaR ``combination'' weights. Namely, as in \cite{giacominikomunjer2005}, for a given VaR level $\alpha_j$ and a forecast origin $t\geq N$, the coefficients $c_{i,j,t}$ used for combining VaR forecasts at time $t$, for $i=0,\ldots,n_{mod}$ (to include the intercept term $c_{0,t}$) can be estimated by minimizing the quantile loss function over a rolling window of fixed size $N$
\begin{align}
    \overline{QL}_{t,N}(\alpha_j,\mathbf{c}_{j,t})&=
\frac{1}{N}\sum_{k=1}^{N}QL(r_{t-k},\widehat{\mathbf{Q}}^{(U,\alpha_j)}_{t-k};\alpha_j)\\ \nonumber
&=
\frac{1}{N}\sum_{k=1}^{N}
(\alpha_j-I_{t-k,j}) \left(r_{t-k}-
\widehat{\mathbf{X}}^{(U,\alpha_j)}_{t-k}\mathbf{c}_{j,t} \right),
\label{e:comb_objective_matrix}
\end{align}
where $\mathbf{c}_{j,t}=(c_{0,j,t},c_{1,j,t},\ldots,c_{n_{mod},j,t})^{'}$, $\widehat{\mathbf{Q}}^{(U,\alpha_j)}_{t}\equiv \widehat{\mathbf{Q}}^{(U,\alpha_j)}_{t:t}$ (a vector of size $1 \times n_{mod}$, from row $t$ in matrix $\widehat{\mathbf{Q}}^{(U,\alpha_j)}_{1:(N+H)}$),
$\widehat{\mathbf{X}}^{(U,\alpha_j)}_{t}=\left[ 1\,\, \widehat{\mathbf{Q}}^{(U,\alpha_j)}_{t}\right]$ (a vector of size $ 1\times (n_{mod}+1)$),
$I_{t,j}=I(r_t<\widehat{\mathbf{X}}^{(U,\alpha_j)}_{t}\mathbf{c}_{j,t})$, for each quantile level $\alpha_j$, $j=1,\ldots, M$. Analytically, the estimated coefficients for combining one-step-ahead VaR forecasts with origin at time $N+h$ are given by
\begin{equation}
\widehat{\mathbf{c}}_{j,N+h}=\underset{\mathbf{c}_{j,N+h}}{ \arg \min}\,\, \overline{QL}_{N+h,N}(\alpha_j,\mathbf{c}_{j,N+h}).
\label{e:combpred_est}
\end{equation}

Combined one-step-ahead VaR forecasts are then finally computed through substituting the fitted $\widehat{\mathbf{c}}_{j,N+h}$ coefficients in Equation (\ref{e:combpred})

\begin{equation}
\widehat{Q}^{(C,\alpha_j)}_{N+h}=\widehat{\mathbf{X}}^{(U,\alpha_j)}_{N+h}\widehat{\mathbf{c}}_{j,N+h}, \qquad \textrm{for } h=1,\ldots,H.
\label{e:combpred_q_matrix}
\end{equation}

%where $\mathbf{c}_{j,t}=(c_{0,j,t},\ldots, c_{n_{mod},j,t})'$ and

%(with a rolling window of fixed size $N$) to produce the combined quantile estimator. Then, Equation (\ref{e:combpred_est}) is used to produced the quantile combination weights.

%\begin{equation}
%\bar{Q}^{(\alpha_j)}_{t}=c_{0,j}+\sum_{i=1}^{n_{mod}}c_{i,j}\widehat{Q}^{(\alpha_j)}_{t,i},
%\label{e:combpred}
%\end{equation}

Step 1 of the FC-WQ generates an output matrix of combined quantile predictors $\widehat{\mathbf{Q}}^{(C)}_{1:T}$ of size $T\times M$. Each column in this matrix is produced from combining conditional quantile estimates/forecasts from $n_{mod}$ models. Namely, the values in column $M$ and rows from $N+1$ to $N+H$ in the matrix $\widehat{\mathbf{Q}}^{(C)}_{1:T}$ contain the 2.5\% target level one-step-ahead quantile forecasts $\widehat{\mathbf{Q}}^{(C,\alpha_M)}_{(N+1):(N+H)}$, noting $\alpha_M=\alpha=2.5\%$.

Rows from 1 to $N$ are computed by estimating the combination weights on the in-sample quantile estimates from the $n_{mod}$ candidate models and then using these weights to combine the time series of in-sample quantile estimates. %In-sample estimates are gradually (in a rolling window manner) replaced with out-of-sample quantile forecasts (with the combined results saved in rows from $N+1$ to $T=N+H$).
The same weights, based on in-sample VaR estimates from time $1$ to $N$, are used to generate row $N+1$ of the matrix. Row $N+2$ is then generated using weights estimated on the $N\times n_{mod}$ time series composed of in-sample quantile estimates from time $2$ to $N$ (for the first $N-1$ rows), and by the formerly generated vector of quantile forecasts at time $N+1$ (for the last row). This procedure is iterated, in a rolling window fashion, until the end of the available sample $T$. A detailed step-by-step implementation description is shown in the Algorithm 1 presented after the illustration of the estimation step 2.

\noindent \textbf{Estimation step 2}:

In the second stage of our approach, we predict the conditional ES at time $N+h$ as an affine function of the elements of $\widehat{\mathbf{Q}}^{(C)}_{N+h}$ (row $N+h$ in matrix $\widehat{\mathbf{Q}}^{(C)}_{1:T}$). The only unknown parameters in Equation (\ref{e:insES}) for calculating the $\text{ES}^{(\text{FC-WQ})}$ estimator are $(w_0,a,b)$. Conditioning on first stage produced VaR series $\widehat{\mathbf{Q}}^{(C)}_{t}$ (a vector of size $1 \times M$, row $t$ in matrix $\widehat{\mathbf{Q}}^{(C)}_{1:T}$) and letting $\boldsymbol{\theta_t}=({w}_{0,t},{a}_{t},{b}_{t})'$, the values of the coefficients used for generating the ES forecast at time $t$ can be estimated by minimizing wrt $\boldsymbol{\theta}_t$ the strictly consistent scoring function:
%\begin{equation*}
%\boldsymbol{\theta}_t
%=\underset{\boldsymbol{\theta}_t}{\arg\min}\,
% \left( \alpha, \boldsymbol{\theta}_t |\mathbf{c}_t \right),
%\end{equation*}
%where
\begin{equation*}
 \bar{S}_{t,N}\left(\alpha,\boldsymbol{\theta}_t|\mathbf{c}_t\right) =\frac{1}{N}\sum_{k=1}^{N} S_{t-k}\left(\alpha, \boldsymbol{\theta}_t |\mathbf{c}_t \right),
\end{equation*}
where
\begin{equation*}
S_t\left(\alpha, \boldsymbol{\theta}_t |\mathbf{c}_t \right)=
-\text{log} \left( \frac{\alpha-1}{\widehat{ES}_{t}} \right) - {\frac{\left(r_t-\widehat{Q}^{(C,\alpha)}_{t}\right)\left(\alpha-I\left(r_{t}\leq \widehat{Q}^{(C,\alpha)}_{t}\right)\right)}{\alpha \,\widehat{ES}_{t}}},
\end{equation*}
and
\begin{equation}
      \widehat{ES}_{t}=\widehat{\mathbf{X}}^{*}_{t}{\mathbf{w}}_{t},
\end{equation}
%\overline{\mathbf{Q}}_{t}=\hat{\mathbf{X}}^{(U,\alpha)}_t \hat{\mathbf{c}}_{M,t}, \,\qquad\,
with $\widehat{\mathbf{X}}^{*}_{t}=[1\,\, \widehat{\mathbf{Q}}^{(C)}_{t}]$ (a vector of size $1 \times (M+1)$) and $\mathbf{w}_t=(w_{0,t},w_{1,t},\ldots,w_{M,t})'$. $\widehat{Q}^{(C,\alpha)}_{t}\equiv\widehat{Q}^{(C,\alpha_M)}_{t}$ is the quantile input at the target $\alpha=\alpha_M=2.5\%$ quantile level and produced from the estimation step 1 (row $t$, column $M$ in matrix $\widehat{\mathbf{Q}}^{(C)}_{1:T}$).
%and $\hat{\mathbf{w}}_t$ denotes its estimate.

%where $S_t(.)$ is defined as in Equation (\ref{es_caviar_log_score}) and $\text{ES}_t$ follows the specification in Equation (\ref{e:insES}).
Namely, the estimated coefficients for weighting the $M$ one-step-ahead combined VaR forecasts for time $N+h$ are obtained as
\begin{equation}
\widehat{\boldsymbol{\theta}}_{N+h}=\underset{\boldsymbol{\theta}_{N+h}}{ \arg \min}\,\, \bar{S}_{N+h,N}  \left( \alpha, \boldsymbol{\theta}_{N+h} |\mathbf{c}_{N+h} \right),
\label{e:es_pred_est}
\end{equation}
where $\widehat{\boldsymbol{\theta}}_{N+h}=( \widehat{w}_{0,N+h},\widehat{a}_{N+h}, \widehat{b}_{N+h} )$. Minimization of the above AL log-score was implemented using the Quasi-Newton optimizer implemented in the Matlab \emph{fminunc} function.

The estimated parameters $\widehat{a}_{N+H}$ and $\widehat{b}_{N+H}$ are plugged in the Beta lag function (\ref{e:betalag}) to produce $\widehat{w}_{1,N+h},\ldots,\widehat{w}_{M,N+h}$. Therefore, the fitted coefficients
\begin{equation}
\widehat{\mathbf{w}}_{N+h}=(\widehat{w}_{0,N+h},\widehat{w}_{1,N+h},\ldots,\widehat{w}_{M,N+h})'
\label{e:es_w_est}
\end{equation}
are then employed in the weighted quantile framework as below to produce the $h$-th one-step-ahead ES forecast at the target $\alpha=2.5\%$ quantile level

\begin{equation}
\widehat{ES}^{(\text{FC-WQ})}_{N+h}=\widehat{\mathbf{X}}^{*}_{N+h}\widehat{\mathbf{w}}_{N+h}, \qquad \textrm{for } h=1,\ldots,H.
\label{e:combpred_es_matrix}
\end{equation}

%\begin{eqnarray*}
%S_t(r_t, ES_t;w_0,a,b|\mathbf{Q}_t) &=& (I_t -\alpha)G_1(Q_{t}) - I_tG_1(r_t) +  G_2(ES_t)\left(ES_t-Q_t + \frac{I_t}{\alpha}(Q_t-r_t)\right)
%\nonumber
%\\
%                      &-& H(ES_t) + a(r_t) \, ,
%                      \label{e:estscore}
%\end{eqnarray*}
% One-step-ahead forecasts of $\text{ES}_t$ can then be easily computed by replacing estimated in-sample quantiles, on the RHS of (\ref{e:insES}), by their out-of-sample forecasts obtained from the associated quantile forecasts $\widehat{\mathbf{Q}}_{(N+1)}$. Formally, the ES predictor at time $N+1$ (\cred{a similar procedure can be applied to all the following forecasting steps}), conditional on in-sample information available at time $N$, is obtained as
% \begin{equation}
% \widehat{ES}^{(\text{FC-WQ})}_{N+1}= \hat{w}_{0,N+1}+\sum_{j=1}^{M} \hat{w}_{j,N+1} \hat{Q}_{N+1}^{(\alpha_j)} \, ,
% \label{e:outsES}
% \end{equation}
% %where the subscript $N$ in $w_{i,N}$ indicates that the weight function is
% where the $\hat{w}_{j,N+1}$ weights depend on the fitted values of $\hat{a}_{N+1}$ and  $\hat{b}_{N+1}$, estimated using information up to time $N$.
%\cred{What about reporting this (useful) description in an Appendix? Yes, this is a great idea.}
For the sake of clarity, the detailed step by step description of the outlined forecasting algorithm is presented in Algorithm 1.

\begin{algorithm}
    \caption{Forecast Combination and Weighted Quantile Algorithm}
  \begin{algorithmic}[1]
    % \REQUIRE Decomposition of signal $x$
    \INPUT In-sample quantile estimates $\widehat{\mathbf{Q}}_{1:N, i}^{(\alpha_j)}$ ($N \times 1$, $i=1,\ldots,n_{mod}$, $j=1,\ldots,M$); one-step-ahead quantile forecasts $\widehat{\mathbf{Q}}_{N+1:N+h, i}^{(\alpha_j)}$ ($H \times 1$, $i=1,\ldots,n_{mod}$, $j=1,\ldots,M$, $h=1,\dots,H$).
    \OUTPUT  One-step-ahead combined VaR forecast $\widehat{Q}_{N+h}^{(C,\alpha_j)}$ ($j=1,\ldots,M$); target 2.5\% ES one-step-ahead forecast $\widehat{ES}^{(\text{FC-WQ})}_{N+h}$, $h=1,\dots,H$.
    %\STATE \textbf{Initialization} $R^{(0)} = x$
    \FOR{$h = 1,\dots, H$}
      \STATE Employ the in-sample quantile estimates $\widehat{\mathbf{Q}}_{h:N, i}^{(\alpha_j)}$ ($i=1,\ldots,n_{mod}$, $j=1,\ldots,M$), one-step-ahead quantile forecasts $\widehat{\mathbf{Q}}_{N+1:N+h-1, i}^{(\alpha_j)}$ ($i=1,\ldots,n_{mod}$, $j=1,\ldots,M$)  and Equation (\ref{e:combpred_est}) to estimate the quantile combination weights $\widehat{\mathbf{c}}_{j,N+h}$
        \COMMENT{In iteration 1, only in-sample quantile estimates $\widehat{\mathbf{Q}}_{1:N, i}^{(\alpha_j)}$ are used};
      %\STATE Employing Equation (\ref{e:combpred}) and estimated weights $\hat{\mathbf{c}}_{j,N+h}$, for $j=1,\ldots,M$, combine the in-sample quantiles $\mathbf{Q}_{1:N,i}^{(\alpha_j)}$ ($i=1,\ldots,n_{mod}$) into a combined quantile predictor ${\mathbf{Q}}_{1:N}^{C,(\alpha_j)}$

      \STATE Incorporate Equation (\ref{e:combpred_q_matrix}), estimated weight $\widehat{\mathbf{c}}_{j,N+h}$, and $h$-th one-step-ahead quantile forecasts $\widehat{Q}_{N+h, i}^{(\alpha_j)}$ ($i=1,\ldots,n_{mod}$, $j=1,\ldots,M$) to produce the combined $h$-th one-step-ahead quantile (VaR) forecast $\widehat{Q}_{N+h}^{(C,\alpha_j)}$. Here, $\widehat{Q}_{N+h}^{(C,\alpha_M)}= \widehat{Q}_{N+h}^{(C,\alpha)}$ is the target 2.5\% quantile level $h$-th one-step-ahead combined VaR forecast;

      \STATE Use the combined quantile estimators $\widehat{\mathbf{Q}}_{h:N+h-1}^{(C)}$ (calculated from the in-sample quantiles $\widehat{\mathbf{Q}}_{h:N, i}^{(\alpha_j)}$ ($i=1,\ldots,n_{mod}$, $j=1,\ldots,M$),  one-step-ahead quantile forecasts $\widehat{\mathbf{Q}}_{N+1:N+h-1, i}^{(\alpha_j)}$ ($i=1,\ldots,n_{mod}$, $j=1,\ldots,M$) and estimated combination weights $\hat{\mathbf{c}}_{j,N+h}$), to estimate the WQ Beta weights $\widehat{\mathbf{w}}_{t+h}$ using Equation (\ref{e:es_w_est});
      \COMMENT{In our study out-of-sample size $H$ is greater than the in-sample size $N$, thus when $h > N$ in-sample quantiles are all replaced by one-step-ahead quantile forecasts when calculating $\widehat{\mathbf{Q}}_{h:N+h-1}^{(C)}$};

      \STATE Employ Equation (\ref{e:combpred_es_matrix}), estimated WQ Beta weights  $\widehat{\mathbf{w}}_{t+h}$ and combined one-step-ahead quantile forecasts $\widehat{\mathbf{Q}}_{N+h}^{(C)}$, to produce the target 2.5\% ES one-step-ahead forecast $\widehat{ES}^{(\text{FC-WQ})}_{N+h}$.

    \ENDFOR
  \end{algorithmic}
\end{algorithm}

{\centering
\section{The model universe}\label{model_universe}
\par
}
\noindent

Various types of models, such as parametric and semi-parametric VaR models, are selected as candidate models in the model universe. All these models, except for the CAViaR \citep{caviar}, can be also used to generate ES forecasts. These ES forecasts will not be of interest for the implementation of the FC-WQ procedure but will be later used as benchmarks for ES forecasts comparison. In total, $n_{mod}=8$ different models are selected with details shown below.

\noindent \textbf{GJR-GARCH-t (parametric)}: the GJR-GARCH model, proposed by \cite{glosten1993relation}, extends the parametric GARCH model to capture the well-known leverage effect (negative returns at time $t-1$ have a larger impact on the volatility at time $t$ than positive returns). In addition, to capture the fat-tail property of financial returns, the Student-t distribution is employed.

\noindent \textbf{EGARCH-t (parametric)}: as another commonly used parametric model, the EGARCH model \citep{nelson1991conditional} is also included as a candidate model. The EGARCH model does not require any positivity restriction on the parameters, since its volatility equation is on log-variance instead of variance itself. Thus the positivity of the variance is automatically satisfied, which is an important advantage of the framework. Again, the Student-t distribution is used to describe the potential leptokurtosis of the conditional return distribution.

\noindent \textbf{POT-GJR-GARCH-t (semi-parametric)}: we also consider some semi-parametric models as part of the model universe. The peaks-over-threshold (POT)-GJR-GARCH-t combines the GJR-GARCH-t model with an extreme value theory \citep{mcneil2000estimation} approach to the fitting of the tail properties of the error distribution. Namely, this is accomplished by applying the POT approach to the returns standardized by the GJR-GARCH-t estimated volatility; see \cite{gilli2006application} for details.

\noindent \textbf{POT-EGARCH-t (semi-parametric)}: this approach is similar to the above described POT-GJR-GARCH-t with the difference that the POT method is applied to returns standardized by the EGARCH-t estimated volatility.

\noindent \textbf{GJR-GARCH-t-HS (semi-parametric)}: a semi-parametric filtered historical simulation approach, which could potentially produce improved tail risk forecasting results, is also included. The series of in-sample conditional variance ($\hat{\sigma}_t$) is estimated based on the fitted GJR-GARCH-t model. The error quantiles and tail expectations are then estimated by computing the relevant sample quantiles $(\hat{q}^{(\alpha)})$ and tail averages $(\hat{c}^{(\alpha)})$ of standardized returns $r_t / \hat{\sigma}_t$. Finally, level-$\alpha$ VaR and ES forecasts are obtained by multiplying $\hat{q}^{(\alpha)}$ and $\hat{c}^{(\alpha)}$, respectively, by the forecast $\hat{\sigma}_{N+1}$ from the fitted GJR-GARCH-t model.

\noindent \textbf{EGARCH-t-HS (semi-parametric)}: this approach employs a similar procedure as the GJR-GARCH-t-HS, by replacing GJR-GARCH-t with EGARCH-t for volatility estimation and forecasting.\\

We would like to emphasise that, for the above models which rely on parametric volatility estimates and forecasts, we do not need to re-estimate the model for each trial quantile level $\alpha_j$; $j=1,\ldots, M$, regarding the VaR and ES calculation. Differently, the remaining two models rely on direct modelling of VaR dynamics. Therefore, they need to be re-estimated for each trial quantile level $\alpha_j$. \\

\noindent \textbf{CAViaR-AS (semi-parametric)}: the CAViaR models proposed by \cite{caviar} are estimated using quantile regression. Although the CAViaR framework does not directly produce ES estimates and forecasts, it can be used without any issues in our FC-WQ approach which only requires the quantile estimates and forecasts from individual models.

Since the CAViaR model is used as the quantile estimation model in the WQ framework of \cite{storti2020nonparametric}, we have deliberately chosen the CAViaR model as part of the model universe, to facilitate the direct comparison of the proposed FC-WQ approach and the WQ approach in the empirical section.

The CAViaR with asymmetric slope (CAViaR-AS) framework which aims to capture the leverage effect is employed:

\begin{equation} \label{caviar_model}
Q_{t}= \beta_0+ \beta_1 Q_{t-1} +(\beta_{2}I_{[r_{t-1} \ge 0]}+\beta_{3}I_{[r_{t-1}<0]}) |r_{t-1}| .
\end{equation}

As documented by \cite{caviar}, solutions to the optimization of the quantile loss objective function can be heavily dependent on the chosen initial values. To account for this issue, we adopt a multi-start optimization procedure inspired by that suggested in \cite{caviar}.

\noindent \textbf{CARE-AS (semi-parametric)}: a different approach to joint estimation of VaR and ES is based on the theory of expectiles. The concept of expectile is closely related to the concept of quantile. The $\tau$ level expectile $\mu_{\tau}$, as defined by \cite{aigneretal1976}, can be estimated through minimizing the following Asymmetric Least Squares (ALS) criterion \citep{newey1987}:
\begin{equation}\label{als_equation}
\sum_{t=1}^{N} |\tau-I(r_t<\mu_{\tau})|(r_t-\mu_{\tau})^2  \, ,
\end{equation}
no distributional assumption is required to estimate $\mu_{\tau}$ here.

\cite{tayl2008} proposes a class of semi-parametric models for VaR and ES forecasting, called Conditional Autoregressive Expectile (CARE) models, with a similar form to the CAViaR model. Under the CARE framework, the lagged returns drive the expectiles and model parameters are estimated via minimizing an ALS criterion.

To select the appropriate expectile levels for VaR and ES estimation, implementation of CARE type models requires a grid search process between 0 and the target quantile level $2.5\%$, based on the optimization of the violation rate (VRate, the percentage of returns exceeding VaR estimates). The size of the expectile level grid search is selected as 100 in our paper.

The CARE with asymmetric slope (CARE-AS) specification as below is included in the model universe, where the expectile responds asymmetrically to positive and negative returns:

\noindent
\begin{eqnarray}\label{care_x_as_equation}
\mu_{t;\tau}=\beta_{0;\tau}+\beta_{1;\tau}\mu_{t-1;\tau}+(\beta_{2;\tau}I_{[r_{t-1}\ge 0]}+\beta_{3;\tau}I_{[r_{t-1}<0]}) |r_{t-1}| \, .
\end{eqnarray}

{\centering
\section{Empirical study}\label{data_empirical_section}
\par
}
\subsection{Data and empirical study design}\label{data_section}
The daily data, including open, high, low and closing prices, are downloaded from Thomson Reuters Tick History and cover the period from the beginning of 2000 to the end of 2015. The closing price is employed to calculate the daily return $r_t$. Data are collected for six market indices: S\&P500 (US), Hang Seng (Hong Kong), FTSE 100 (UK), DAX (Germany), SMI (Swiss) and ASX200 (Australia).
%The starting date of the data is chosen as early as 2000.
%$\alpha= 2.5\%$ is employed for both one day ahead VaR and ES forecasting study for the 7 indices.

As described in Section \ref{two_step_estimation}, a rolling window with fixed in-sample size is employed for estimation and to produce each one-step-ahead forecast in the forecasting period. Table \ref{t:summ_var_fore} reports the in-sample size for each series, which differs due to different non-trading days occurring in each market.

The forecasting study incorporates a 8 year out-of-sample period, with the start date of the out-of-sample chosen as January 2008 (to include the 2008 GFC as part of the out-of-sample period) and out-of-sample size $H$ as 2000. Therefore, the end of the forecasting period is around the end of 2015, with small differences among different markets due to calendar effects.

Employing the proposed FC-WQ framework, both daily one-step-ahead VaR and ES forecasts are produced for the returns on the six indices. VaR forecasts are produced for the whole range of selected trial quantile values defined in Section \ref{two_step_estimation} while, for ES forecasting, $\alpha= 2.5\%$ is chosen as target level, as recommended by Basel Committee on Banking Supervision (2019).

For comparison, VaR and ES forecasts are also generated from each individual model included in the model universe, as presented in Section \ref{model_universe}. Since the CAViaR-AS model cannot directly produce the ES forecasts, the ES-CAViaR models of \cite{tayl2017} are also included in the ES study, again employing the CAViaR-AS models as the specification of the quantile regression component. Two VaR to ES relationships, additive and multiplicative, are employed for the ES-CAViaR framework. We name the models as ES-CAViaR-Add-AS and ES-CAViaR-Mult-AS, respectively. Then, to assist the optimization in the estimation of ES-CAViaR models, the initial values of the parameters of the ES component are also selected by means of an additional random sampling procedure, following \cite{tayl2017}.

The WQ approach of \cite{storti2020nonparametric} using CAViaR-AS as the quantile estimation model is also included for comparison.

\subsection{VaR forecasts evaluation}\label{quantile_fore_section}

One-step-ahead forecasts of VaR and ES are generated for each day in the forecasting period for each data series. This section focuses on the evaluation of VaR forecasts. For brevity, we only report results for the $2.5\%$ quantile level. However, we would like to emphasize that the adopted quantile forecast combination approach allows to reduce the impact of model uncertainty, potentially improving the quantile estimation and forecasting accuracy, for each trial quantile level, i.e., $\alpha_1, \alpha_2, \ldots, \alpha_M$. This is expected to positively affect the second step ES estimation and forecast (details to be shown in the following section).
%For this section, only the target 2.5\% quantile loss FC-WQ results are compared.

First, the VaR violation rate (VRate) is employed to initially assess VaR forecasting accuracy. VRate is simply the proportion of returns that exceed the forecasted VaR in the forecasting period, as in Equation (\ref{varvrate_equation})
\begin{equation}\label{varvrate_equation}
\text{VRate}= \frac{1}{H} \sum_{t=N+1}^{N+H} I(r_t<\text{VaR}_t)\, ,
\end{equation}
where $N$ is the in-sample size and $H=2000$ is the out-of-sample size. Models with a VRate closest to the nominal quantile level $\alpha= 2.5\%$ are preferred, or equivalently $\frac{\text{VRate}}{\alpha}$ closest to 1.

Table \ref{t:summ_var_fore} summarizes the $\frac{\text{VRate}}{\alpha}$ (the closer to 1 the better) at the 2.5\% quantiles over the six indices for all competing models. The ``MAD'' column shows the Mean Absolute Deviation, employing 2.5\% as the target VRate, across the six indices. Box indicates the best model, while dashed box indicates the 2nd best model.

Overall, the proposed FC-WQ frameworks produce favourable VRate results, compared with the other 8 competing individual models in the model universe. The ``MAD'' value from FC-WQ is 0.0028 which is the smallest, followed by CARE-AS.

Here, we would like to mention that only the first step (quantile forecast combination) in the proposed FC-WQ framework would affect the VaR forecasting performance. The second step (weighted combined quantile of each trial quantile level) will determine the ES forecasting performance which will be presented in the following section.

% Here, we would like to mention that the choice of $M=3$ or $M=5$ would not affect the target level quantile forecasts. The differences between the slightly different performance between FC-WQ-3 and FC-WQ-5 are mainly due to the differences produced from the individual estimation procedure, such as CAViaR-AS. \cred{or we might simply present one FC-WQ results only, similar to our WQ paper.}

\begin{table}[hbt!]
\begin{center}
\caption{\label{t:summ_var_fore} \small $\frac{\text{VRate}}{\alpha}$ across the six markets.}\tabcolsep=10pt
\tiny
\begin{tabular}{lcccccc|c} \hline

Model&S\&P500&HangSeng&FTSE&DAX&SMI&ASX200&MAD\\\hline
GJR-GARCH-t&1.68&1.26&1.50&1.68&1.36&1.58&0.0128\\
EGARCH-t&1.62&1.24&1.50&1.64&1.48&1.50&0.0124\\
POT-GJR-GARCH-t&1.30&1.08&1.06&1.26&1.12&1.14&0.0040\\
POT-EGARCH-t&1.26&1.10&1.08&1.18&1.18&1.10&0.0038\\
GJR-GARCH-t-HS&1.30&1.06&1.04&1.26&1.10&1.10&0.0036\\
EGARCH-t-HS&1.26&1.10&1.06&1.16&1.16&1.10&0.0035\\

CAViaR-AS&	 1.18 &	 1.00& 	 1.08 &	 1.28 	& 1.28 	& 1.10 & 0.0038  \\

CARE-AS&1.08&1.00&1.16&1.24&1.24&0.98&\dbox{0.0031}\\

% ES-CAViaR-Add-AS&1.22&0.9&1.10&1.28&1.22&1.02&5.17&0.0039\\
% ES-CAViaR-Mult-AS&1.16&0.96&1.14&1.30&1.24&1.10&5.00&0.0041\\

% FC-WQ-3&1.06&1.02&1.04&1.20&1.20&0.86&\dbox{4.00}&\fbox{0.0028}\\
FC-WQ&1.04&0.98&1.00&1.18&1.32&0.90&\fbox{0.0028}\\
\hline
Out-of-sample $H$ &2000&2000&2000&2000&2000&2000&\\
In-sample $N$&1905&1890&1943&1936&1930&1871&\\
\hline
\end{tabular}
\end{center}
\emph{Note}: \small  Box indicates the favoured model and dashed box indicates the 2nd ranked model based on the average MAD.
\end{table}

The average value of the quantile loss over the out-of-sample period is then used to compare the VaR forecast accuracy of competing models. This choice is motivated considering that the standard quantile loss function is strictly consistent, i.e., the expected loss is a minimum at the true quantile series. The quantile loss function is the one that is employed to optimize the quantile forecast combination weights, as described in Section \ref{two_step_estimation}. Thus, the most accurate VaR forecasting model is expected produce the minimized aggregated quantile loss function, given as in Equation (\ref{q_loss_score}):

\begin{equation}\label{q_loss_score}
\sum_{t=N+1}^{N+H}(\alpha-I(r_t<Q_t))(r_t-Q_t)  \,\, ,
\end{equation}
where $N$ is the in-sample size and $H=2000$ is the out-of-sample size. $\widehat{Q}_{N+1},\ldots,\widehat{Q}_{N+H}$ is a series of quantile forecasts at level $\alpha=2.5\%$ for the observations $r_{N+1},\ldots,r_{N+H}$.

The values of the out-of-sample quantile loss are presented in Table \ref{t:25_quantileloss}. The average loss is included in the ``Avg Loss'' column. Still, box indicates the favoured model and dashed box indicates the 2nd ranked model based on the average loss.

Based on the quantile loss results, we can see that the proposed FC-WQ framework is characterized by very competitive performances, with the smallest average quantile loss value 164.2. The GJR-GARCH-t and EGARCH-t are in general least preferred, with the average loss as 166.6 and 166.4 respectively. Although GJR-GARCH-t and EGARCH-t are included in the forecast combination process, the FC-WQ is still capable of producing competitive quantile forecasting results, which lends evidence on the combination weights estimation scheme, as described in Section \ref{two_step_estimation}.

\begin{table}[hbt!]
\begin{center}
\caption{\label{t:25_quantileloss} \small 2.5\% quantile loss function values across the markets.}\tabcolsep=10pt
\tiny
\begin{tabular}{lcccccc|c} \hline

Model&S\&P500&HangSeng&FTSE&DAX&SMI&ASX200&Avg Loss\\ \hline
GJR-GARCH-t&162.9&196.4&156.4&182.9&159.4&141.7&166.6\\
EGARCH-t&166.9&194.9&155.2&181.6&159.3&140.3&166.4\\
POT-GJR-GARCH-t&161.1&195.0&154.5&180.7&159.3&139.8&165.1\\
POT-EGARCH-t&163.8&193.7&153.0&179.5&157.5&138.4&164.3\\
GJR-GARCH-t-HS&161.0&194.9&154.5&180.7&159.3&139.8&165.0\\
EGARCH-t-HS&163.7&193.6&153.0&179.5&157.5&138.4&\dbox{164.3}\\

CAViaR-AS	& 167.7 &	 190.6 	 & 153.1 &	 179.2& 	 159.0 	 &139.9 &	 164.9 \\

CARE-AS&168.3&189.7&154.7&180.8&163.6&142.4&166.6\\

% ES-CAViaR-Add-AS&166.5&190.4&153.2&179.2&158.6&140.6&164.8&4.50\\
% ES-CAViaR-Mult-AS&165.2&192.4&152.7&179.8&158.4&139.2&164.6&\dbox{3.83}\\

%FC-WQ-3&160.5&191.6&152.4&181.3&160.0&139.2&\fbox{164.1}&4.83\\
FC-WQ&160.7&191.4&155.1&180.2&159.3&138.5&\fbox{164.2}\\
\hline
\end{tabular}
\end{center}
\emph{Note}:\small  Box indicates the favoured model and dashed box indicates the 2nd ranked model based on the average loss.
\end{table}

\begin{figure}[htp]
	\centering
	\includegraphics[width=1\textwidth]{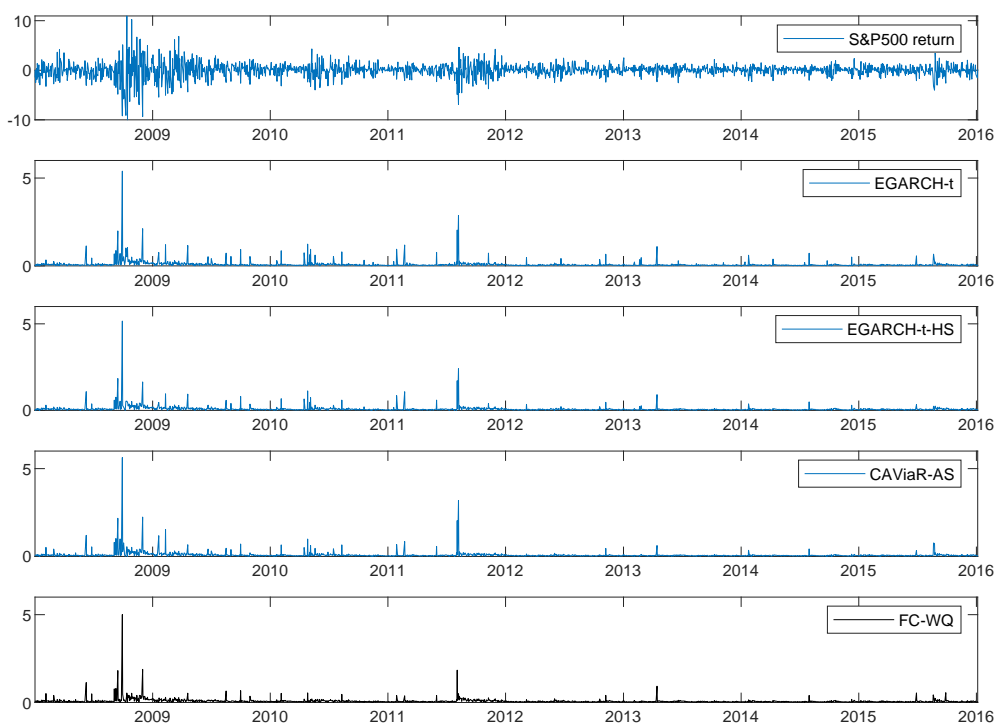}
	\caption{\label{quantile_loss_plot} S\&P500 out-of-sample quantile loss value from EGARCH-t, EGARCH-t-HS, CAViaR-AS and FC-WQ.}
\end{figure}

In addition, for S\&P 500 the quantile loss values for each time step across the whole forecasting period are visualised in Figure \ref{quantile_loss_plot}. Namely,  EGARCH-t, EGARCH-t-HS, CAViaR-AS (used in the WQ approach of \citealp{storti2020nonparametric}) and FC-WQ are compared. The forecast combination is evidently characterized by more ``stabilized'' quantile loss values. For example, between 2009 and 2012, the quantile loss from the forecast combination approach is consistently smaller than that of the competing models, including the EGARCH-t-HS and CAViaR-AS which have good quantile loss performance as shown in Table \ref{t:25_quantileloss}. Therefore, the proposed forecast combination not only allows obtaining an improved predictor via combining different functional forms, i.e., parametric and semi-parametric models, but also produces more robust and stabilized quantile forecasts through the quantile combination process. Such time stability argument is consistent with the aim of forecast combinations that is to account for model uncertainty and to provide a forecasting performance that is optimal (or close to being optimal) and stable across time, which is supported by the results in Table \ref{t:summ_var_fore}, \ref{t:25_quantileloss} and Figure \ref{quantile_loss_plot}.

Lastly, to further assess the validity of the quantile forecast combination, we have employed the VaR calibration tests defined in \cite{pattonetal2019}, with the code developed by the authors. The test employs a MZ type regression of generalized VaR residuals on fitted VaR and lagged generalized residuals. Regression coefficients are fitted by Ordinary Least Squares (OLS) and standard errors are computed by a Newey-West estimator with 20 lags. The p-values of each data set and model produced from the test are presented in Table \ref{t:25_quantile_backtest}. In addition, on the 10\% significance level, the column ``Total'' shows the total number of rejections (p-value less than 10\%) for each model. Regarding the results, CAViaR-AS receives the least number of rejections and is closely followed by several models, including the proposed FC-WQ, which are rejected twice. The GJR-GARCH-t and EGARCH-t models are the most frequently rejected by the test.

\begin{table}[hbt!]
\begin{center}
\caption{\label{t:25_quantile_backtest} \small 2.5\% VaR calibration test at the 10\% significance level.}\tabcolsep=10pt
\tiny
\begin{tabular}{lcccccc|c} \hline
Model&S\&P500&HangSeng&FTSE&DAX&SMI&ASX200&Total\\ \hline
GJR-GARCH-t&0.003&0.494&0.010&0.001&0.025&0.002&5\\
EGARCH-t&0.056&0.576&0.018&0.006&0.002&0.024&5\\
POT-GJR-GARCH-t&0.255&0.948&0.750&0.037&0.000&0.000&3\\
POT-EGARCH-t&0.740&0.936&0.757&0.639&0.059&0.000&\dbox{2}\\
GJR-GARCH-t-HS&0.255&0.953&0.807&0.037&0.000&0.000&3\\
EGARCH-t-HS&0.740&0.936&0.838&0.706&0.074&0.000&\dbox{2}\\
CAViaR-AS& 0.816	&0.849&	0.796&	0.501&	0.106&	0.000 & \fbox{1}\\

CARE-AS&0.910&0.610&0.734&0.605&0.097&0.000&\dbox{2}\\

% ES-CAViaR-Add-AS&0.758&0.744&0.776&0.492&0.022&0.000&\dbox{2}\\
% ES-CAViaR-Mult-AS&0.746&0.690&0.566&0.416&0.127&0.000&\fbox{1}\\

% FC-WQ-3&0.41&0.906&0.883&0.643&0.052&0&1\\
FC-WQ&0.932&0.682&0.873&0.738&0.069&0.000&\dbox{2}\\
\hline
\end{tabular}
\end{center}
\emph{Note}: \small  Box indicates the favoured model and dashed box indicates the 2nd ranked model based on the total number of rejections on the 10\% significance level. p-values are presented for each index and each model.
\end{table}

\subsection{ES forecasts evaluation}\label{es_fore_section}

We remind that a key feature of the proposed framework is that, for each trial quantile level, the combined VaR predictor can potentially have a different structure in terms of included models and assigned weights. This flexibility can only expected to be beneficial for the ES forecasting performance. Also, no specific assumptions are formulated on relationship linking VaR and ES but, consistently with its theoretical definition, ES is computed as a weighted average of combined VaR forecasts. This design naturally yields a combined ES predictor that has been purged of the impact of model uncertainty. Due to these features, it is expected that,  compared to single forecasting models, the FC-WQ framework could be characterized by an improved and more stable ES forecasting performance. Aim of this section is to provide empirical evidence supporting this hypothesis.

In the FC-WQ framework, we consider $M=3$ and $M=5$ quantile trial levels for the weighted quantile process, with the corresponding framework named as FC-WQ-$M$, $M=3,5$. \cite{storti2020nonparametric} show that the weighted quantile framework is not sensitive to the choice of $M$, meaning a small value of $M$, i.e., 3 or 5, can be chosen in real data applications to reduce the computation requirement.

The WQ approach employing CAViaR-AS for quantile estimation is also included for comparison. We name this model as CAViaR-AS-WQ. Again, $M=3$ and $M=5$ quantile trial levels are considered for the CAViaR-AS-WQ model (CAViaR-AS-WQ-3, CAViaR-AS-WQ-5).

In addition, we also test the ES forecasting performance via incorporating the Simple Average of the combined quantile forecasts (FC-SA), as in below Equation (\ref{e:fc_sa}). The value of $M$ is also selected as 3 and 5 respectively.

\begin{equation}
ES^{(\text{FC-SA})}_{t}= \frac{1}{M} \sum_{j=1}^{M}  \widehat{Q}^{(C, \alpha_j)}_{t} \, .
\label{e:fc_sa}
\end{equation}

Lastly, the Simple Average of the CAViaR-AS estimated quantiles (CAViaR-AS-SA) is included, with the value of $M$ chosen as 3 and 5 respectively as well.

To evaluate the FC-WQ framework comprehensively, we assess the ability of the different models under comparison to forecast VaR and ES jointly, employing the joint loss values in Equation (\ref{es_caviar_log_score}). We use this to jointly compare the VaR and ES forecasts from all models, because the AL log-score in Equation (\ref{es_caviar_log_score}) is a strictly consistent scoring function that is jointly minimized by the true VaR and ES series.

\begin{figure}[htp]
	\centering
	\includegraphics[width=1\textwidth]{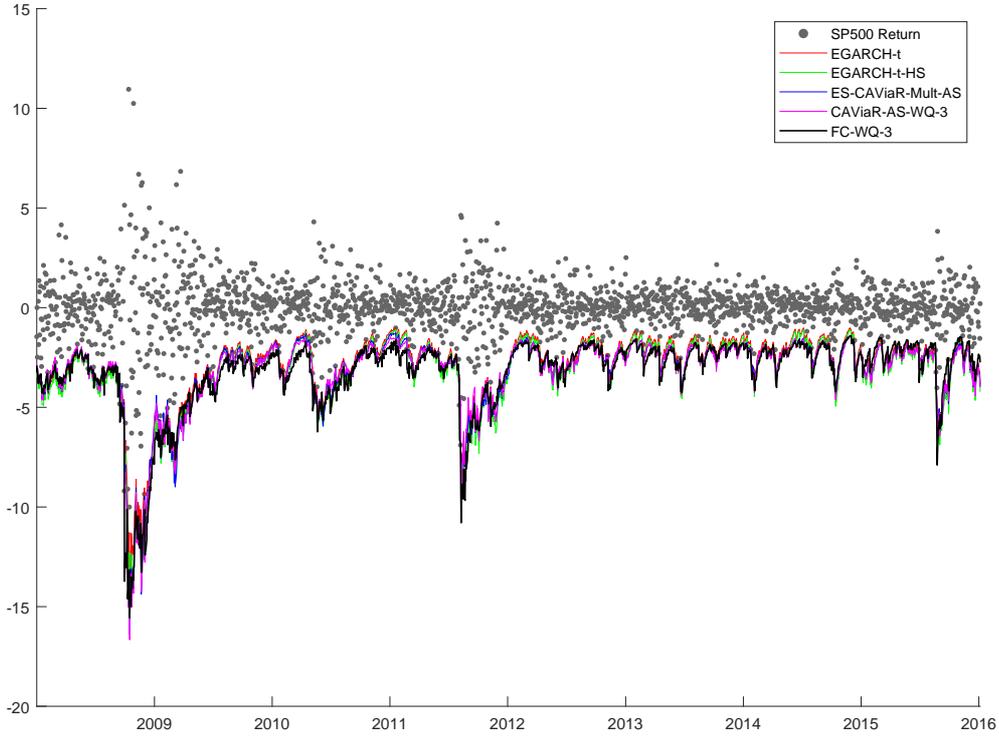}
	\caption{\label{es_forecast_plot} S\&P500 ES forecasts from EGARCH-t, EGARCH-t-HS, ES-CAViaR-Mult-AS, CAViaR-AS-WQ-3 and FC-WQ-3.}
\end{figure}

\begin{figure}[htp]
	\centering
	\includegraphics[width=1\textwidth]{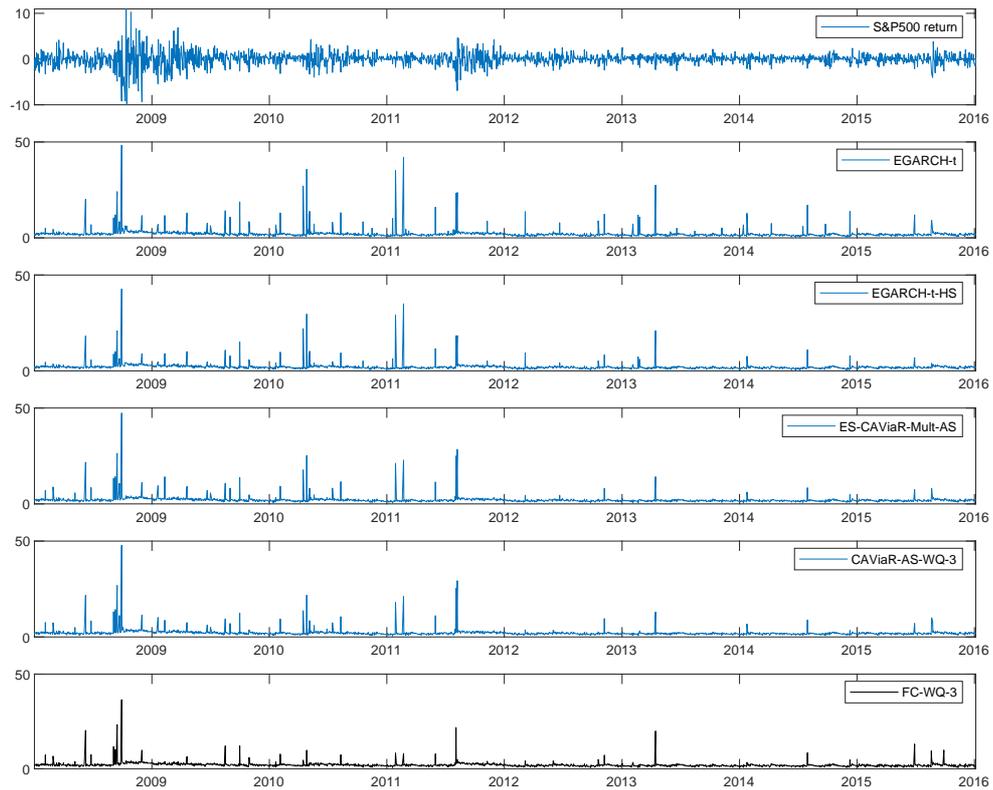}
	\caption{\label{jonit_loss_plot} S\&P500 out-of-sample VaR and ES AL joint loss from EGARCH-t, EGARCH-t-HS, ES-CAViaR-Mult-AS, CAViaR-AS-WQ-3 and FC-WQ-3.}
\end{figure}

First, Figure \ref{es_forecast_plot} shows the S\&P500 ES forecasts from EGARCH-t, EGARCH-t-HS, ES-CAViaR-Mult-AS, CAViaR-AS-WQ-3 and FC-WQ-3. To make a more in-depth comparison of these models, Figure \ref{jonit_loss_plot} presents the S\&P 500 AL joint loss (log-score) values for each time step across the out-of-sample period. In general, we have a consistent story as in the quantile loss plot in Figure \ref{quantile_loss_plot}. The ES forecasts produced from FC-WQ framework are again characterized by more stabilized and smaller joint loss values (i.e., the 2009 to 2012 period) than the ones from the competing individual and CAViaR-AS-WQ models, such as EGARCH-t, EGARCH-t-HS, ES-CAViaR-Mult-AS and CAViaR-AS-WQ-3.

% \begin{figure}[htp]
% 	\centering
% 	\includegraphics[width=1\textwidth]{jonit_loss_plot.eps}
% 	\caption{\label{jonit_loss_plot} Old- S\&P500 out-of-sample VaR and ES AL joint loss from EGARCH-t, EGARCH-t-HS, ES-CAViaR-Mult-AS and FC-WQ-3.}
% \end{figure}

Table \ref{t:25_veloss} reports, for each model and data series, the value of the loss function in Equation (\ref{es_caviar_log_score}) aggregated over the out-of-sample period: $\mathbf{S}= \sum_{t=N+1}^{N+H} S_{t}$, with $H=2000$. In general, the proposed FC-WQ models produce on average the smallest joint loss, i.e, 4257.6 for FC-WQ-3 and 4257.8 for FC-WQ-5. The results of FC-WQ-3 and FC-WQ-5 are quite close to each other, which is consistent with the observations in \cite{storti2020nonparametric} and means the choice of $M=3$ can already produce ES forecast with good accuracy.

Comparing the FC-WQ and CAViaR-AS-WQ results, the FC-WQ produces smaller joint loss values, which lends evidence on the improved quantile and ES forecasts via the proposed quantile forecast combination and weighed quantile approach. The effectiveness of quantile forecast combination is also supported by the findings that the FC-SA models produce smaller joint loss values than the CAViaR-AS-SA models.

The FC-SA framework produces joint loss values which are consistently larger than that from the corresponding FC-WQ, i.e., comparing FC-SA-M to FC-WQ-M ($M=3,5$). Similar findings are observed when comparing CAViaR-AS-SA and CAViaR-AS-WQ. This lends evidence on the effectiveness of employing the Beta weighting scheme on the combined quantile forecasts. Lastly, GJR-GARCH-t and EGARCH-t, which are included in the model universe of the FC-WQ framework, are least preferred, with average loss as 4314.8 and 4315.1 respectively.

\begin{table}[hbt!]
\begin{center}
\caption{\label{t:25_veloss} \small 2.5\% VaR and 2.5\% ES joint loss function values across the markets.}\tabcolsep=10pt
\tiny
\begin{tabular}{lcccccc|c} \hline

Model&S\&P500&HangSeng&FTSE&DAX&SMI&ASX200&Avg Loss\\ \hline

GJR-GARCH-t&4239.2&4601.7&4183.8&4601.4&4253.6&4009.1&4314.8\\
EGARCH-t&4290.6&4586.9&4192.1&4585.2&4245.2&3990.4&4315.1\\
POT-GJR-GARCH-t&4175.3&4587.6&4131.4&4557.7&4215.5&3963.5&4271.8\\
POT-EGARCH-t&4226.2&4575.6&4146.7&4543.2&4185.6&3954.6&4272.0\\
GJR-GARCH-t-HS&4174.1&4587.7&4131.3&4556.6&4216.0&3962.5&4271.3\\
EGARCH-t-HS&4225.1&4575.0&4146.9&4542.8&4185.5&3953.9&4271.5\\
CARE-AS&4276.8&4550.7&4160.6&4514.8&4252.5&4024.9&4296.7\\

ES-CAViaR-Add-AS&4242.3&4551.9&4131.8&4506.8&4192.1&3992.9&4269.6\\
ES-CAViaR-Mult-AS&4242.3&4564.2&4117.8&4509.1&4188.3&3977.4&4266.5\\

CAViaR-AS-SA-3&4265.5&4550.9&4145.5&4519.2&4214.4&3988.1&4280.6\\
CAViaR-AS-SA-5&4275.3&4553.0&4147.5&4525.4&4218.2&3984.9&4284.0\\

CAViaR-AS-WQ-3&4252.6&4549.4&4130.4&4507.7&4195.2&3982.4&4269.6\\
CAViaR-AS-WQ-5&4255.0&4550.9&4135.6&4513.3&4195.2&3979.4&4271.6\\

FC-SA-3&4157.7&4572.8&4138.7&4551.7&4211.5&3966.8&4266.5\\
FC-SA-5&4149.7&4568.1&4182.3&4534.7&4213.1&3946.6&4265.8\\

FC-WQ-3&4152.9&4568.1&4136.5&4537.1&4193.7&3957.5&\fbox{4257.6}\\
FC-WQ-5&4141.9&4565.6&4174.6&4520.8&4198.6&3945.6&\dbox{4257.8}\\

\hline
\end{tabular}
\end{center}
\emph{Note}:\small  Box indicates the favoured model and dashed box indicates the 2nd ranked model based on the average loss.
\end{table}

Lastly, similar to the VaR calibration test, following \cite{pattonetal2019} an ES regression-based calibration test is also conducted with results shown in Table \ref{t:25_es_backtest}. Overall, the observations are similar to that of the VaR calibration test. On the 10\% significance level, the proposed FC-WQ framework together with FC-SA and CARE-AS are least likely to be rejected by the test, compared with other competing models. The CAViaR-AS-WQ and CAViaR-AS-SA models are rejected on 3 markets via the calibration test. The GJR-GARCH-t and EGARCH-t are rejected for all six data sets. However, the proposed FC-WQ framework which includes the GJR-GARCH-t and EGARCH-t in its model universe can still produce competitive ES forecasts via the quantile forecast combination and weighting scheme, which again lends support on its effectiveness.

\begin{table}[hbt!]
\begin{center}
\caption{\label{t:25_es_backtest} \small 2.5\% ES calibration test at the 10\% significance level.}\tabcolsep=10pt
\tiny
\begin{tabular}{lcccccc|c} \hline
Model&S\&P500&HangSeng&FTSE&DAX&SMI&ASX200&Total\\ \hline

GJR-GARCH-t&0.000&0.037&0.000&0.000&0.000&0.000&6\\
EGARCH-t&0.003&0.066&0.000&0.000&0.000&0.001&6\\
POT-GJR-GARCH-t&0.067&0.368&0.135&0.003&0.000&0.000&4\\
POT-EGARCH-t&0.201&0.338&0.161&0.079&0.008&0.000&\dbox{3}\\
GJR-GARCH-t-HS&0.069&0.420&0.175&0.003&0.000&0.000&4\\
EGARCH-t-HS&0.205&0.345&0.207&0.091&0.011&0.000&\dbox{3}\\
CARE-AS&0.508&0.473&0.122&0.180&0.009&0.000&\fbox{2}\\

ES-CAViaR-Add-AS&0.241&0.725&0.123&0.059&0.014&0.000&\dbox{3}\\
ES-CAViaR-Mult-AS&0.298&0.696&0.086&0.055&0.018&0.000&4\\

CAViaR-AS-SA-3&0.285&0.695&0.169&0.059&0.013&0.000&\dbox{3}\\
CAViaR-AS-SA-5&0.292&0.670&0.173&0.058&0.013&0.000&\dbox{3}\\

CAViaR-AS-WQ-3&0.291&0.612&0.137&0.061&0.012&0.000&\dbox{3}\\
CAViaR-AS-WQ-5&0.296&0.579&0.139&0.060&0.012&0.000&\dbox{3}\\

FC-SA-3&0.458&0.541&0.311&0.131&0.007&0.000&\fbox{2}\\
FC-SA-5&0.729&0.506&0.343&0.170&0.004&0.000&\fbox{2}\\

FC-WQ-3&0.459&0.553&0.306&0.119&0.009&0.000&\fbox{2}\\
FC-WQ-5&0.712&0.450&0.344&0.166&0.006&0.000&\fbox{2}\\

\hline
\end{tabular}
\end{center}
\emph{Note}: \small  Box indicates the favoured model and dashed box indicates the 2nd ranked model based on the total number of rejections on the 10\% significance level. p-values are presented for each index and each model.
\end{table}

% \begin{table}[hbt!]
% \begin{center}
% \caption{\label{t:25_quantileloss} \small 2.5\% VaR and 2.5\% ES joint loss function values across the markets.}\tabcolsep=10pt
% \tiny
% \begin{tabular}{lcccccc|cc} \hline

% Model&S\&P500&HangSeng&FTSE&DAX&SMI&ASX200&Avg Loss&Avg Rank\\ \hline

% GJR-GARCH-t&4239.2&4601.7&4183.8&4601.4&4253.6&4009.1&4314.8&12.00\\
% EGARCH-t&4290.6&4586.9&4192.1&4585.2&4245.2&3990.4&4315.1&11.50\\
% POT-GJR-GARCH-t&4175.3&4587.6&4131.4&4557.7&4215.5&3963.5&4271.8&7.50\\
% POT-EGARCH-t&4226.2&4575.6&4146.7&4543.2&4185.6&3954.6&4272.0&6.67\\
% GJR-GARCH-t-HS&4174.1&4587.7&4131.3&4556.6&4216.0&3962.5&4271.3&7.33\\
% EGARCH-t-HS&4225.1&4575.0&4146.9&4542.8&4185.5&3953.9&4271.5&6.00\\
% CARE-AS&4276.8&4550.7&4160.6&4514.8&4252.5&4024.9&4296.7&8.50\\
% FC-WQ-3&4152.9&4568.1&4136.5&4537.1&4193.7&3957.5&\fbox{4257.6}&\dbox{5.17}\\
% FC-WQ-5&4141.9&4565.6&4174.6&4520.8&4198.6&3945.6&\dbox{4257.8}&\fbox{4.83}\\

% Mean&4191.1&4559.8&4136.7&4534.7&4198.1&3949.8&4261.7&\fbox{4.83}\\
% Median&4203.4&4570.0&4131.6&4543.9&4215.9&3946.3&4268.5&6.17\\
% \hline
% ES-CAViaR-Add-AS-MLE&4242.3&4551.9&4131.8&4506.8&4192.1&3992.9&4269.6&5.50\\
% ES-CAViaR-Mult-AS-MLE&4242.3&4564.2&4117.8&4509.1&4188.3&3977.4&4266.5&5.00\\
% \hline
% \end{tabular}
% \end{center}
% \emph{Note}:\small  Box indicates the favoured model and dashed box indicates the 2nd ranked model based on the average loss and rank.
% \end{table}

{\centering
\section{Conclusion}
\label{conclusion}
\par
}

In this paper, in order to reduce the impact of model uncertainty in tail risk forecasting, we propose an innovative framework based on a forecast combination and weighted quantile (FC-WQ) approach that extends the WQ approach in \cite{storti2020nonparametric}.
The first step forecast combination procedure combines quantile forecasts from all VaR models included in the model universe, on a grid of quantile levels. Then, the combined quantiles forecasts are employed as input to a quantile weighting scheme which is used to produce ES forecasts. The coefficients involved in the two steps of the procedure are estimated via optimizing the quantile loss and a strictly consistent joint VaR and ES loss, respectively. The selected model universe consists of parametric and semi-parametric models.

Compared to the VaR \& ES forecast combination approach in \cite{taylor2020forecast}, which combines ``pairs'' of VaR \& ES forecasting from various models, our approach breaks the tie between the VaR and ES model, and only requires VaR forecasts from individual VaR models. Therefore, we implicitly consider a greater variety of functional forms without making the number of parameters explode.

In a comprehensive empirical study, improvements in the out-of-sample forecasting of tail risks, especially ES, are observed, compared to each individual model in the model universe, the simple average approach and the WQ approach of \cite{storti2020nonparametric}. A further advantage of the proposed forecast combination and weighted quantile framework is its attitude to return ``stabilized'' quantile and joint loss values.

The proposed framework can be extended in a number of directions. First, other quantile forecast combination schemes, i.e., considering a multivariate quantile framework in the spirit of \cite{whitetal2008}, could be considered as alternatives. Second, the first step quantile forecast combination approach could be replaced by a cross validation approach which selects one model for each quantile level in the chosen grid. Third, a greater range of models, such as the ones incorporating high frequency based realized measures, could be included in the model universe. All these extensions are currently left for future investigation.

\bibliographystyle{chicago}
\bibliography{bibliography}

\end{document}